# THE WIGNER FUNCTION NEGATIVE VALUE DOMAINS AND ENERGY FUNCTION POLES OF THE HARMONIC OSCILLATOR

**E.E. Perepelkin[a,b,c], B.I. Sadovnikov[a], N.G. Inozemtseva[b,c], E.V. Burlakov[a,b]**

[a] *Faculty of Physics, Lomonosov Moscow State University, Moscow, 119991 Russia*
[b] *Moscow Technical University of Communications and Informatics, Moscow, 123423 Russia*
[c] *Dubna State University, Moscow region, Moscow, 141980 Russia*
*E. Perepelkin e-mail: pevgeny@jinr.ru*

**Abstract**
For a quantum harmonic oscillator an explicit expression that describes the energy distribution as a coordinate function is obtained. The presence of the energy function poles is shown for the quantum system in domains where the Wigner function has negative values.

**Introduction**

The Wigner function [1] is one of the most effective tools for describing quantum systems in phase space. The Wigner function is a quasi-probability density set in phase space. It was constructed in such a way that it can be used to calculate quantum-mechanical averages of observables, similar to how it is done in classical statistical physics, without involving the operators of the corresponding quantities. The Wigner function is widely used in quantum tomography [2], quantum communication and cryptography [3], quantum informatics [4], in signal processing problems [5]. The traditional definition of the Wigner function is given in terms of the density matrix of a quantum system:

$$W(\vec{r},\vec{p},t) = \frac{1}{(2\pi\hbar)^3} \int_{(\infty)} \exp\left(-i\frac{\vec{p}\vec{s}}{\hbar}\right) \left\langle \vec{r}+\frac{\vec{s}}{2} \middle| \hat{\rho}(t) \middle| \vec{r}-\frac{\vec{s}}{2} \right\rangle d^3 s. \qquad (i.1)$$

For a quantum system described by the wave function $\Psi(\vec{r},t)$ or $\tilde{\Psi}(\vec{p},t)$, the integration of the function $W$ over the space of momenta and coordinates gives the correct relations for the functions $|\Psi|^2$ and $|\tilde{\Psi}|^2$, respectively:

$$\int_{(\infty)} W(\vec{r},\vec{p},t) d^3 p = |\Psi(\vec{r},t)|^2, \quad \int_{(\infty)} W(\vec{r},\vec{p},t) d^3 r = |\tilde{\Psi}(\vec{p},t)|^2. \qquad (i.2)$$

This shows that this definition (i.1) is consistent with a natural property inherent in classical distribution functions. However, both estimation for the integral (i.1) and direct calculation of the Wigner functions according to this definition show that it can take negative values. This fact does not allow us to consider it as a rigorous distribution function. The question of the negativity of the quasi-probability density has been discussed in many works [6-9].

As an alternative to the Wigner function for some classes of problems various variants of $P, Q-$ distribution functions are phenomenologically constructed [11-13], which are positive in the phase space, but a property similar to (i.2) no longer holds for such functions. Thus, in the quantum case, the use of any distribution function on the phase space leads to the violation of one or the other requirements of the classical probability theory. The problems regarding the search for positive quasi-distribution functions as well as possible interpretations of negative quasi-probability values remain to be challenging today.

A.A. Vlasov obtained an infinite self-linked chain of equations for the distribution density functions of higher kinematic quantities $f_1(\vec{r},t)$, $f_2(\vec{r},\vec{v},t)$, $f_3(\vec{r},\vec{v},\dot{\vec{v}},t)$,... [16]. Let us consider the first two equations from the infinite self-linked Vlasov equations chain for the probability density distribution functions $f_1(\vec{r},t)$ and $f_2(\vec{r},\vec{v},t)$:

$$\frac{\partial}{\partial t} f_1(\vec{r},t) + \operatorname{div}_r \left[ \langle \vec{v} \rangle (\vec{r},t) f_1(\vec{r},t) \right] = 0, \quad (i.3)$$

$$\frac{\partial}{\partial t} f_2(\vec{r},\vec{v},t) + \operatorname{div}_r \left[ \vec{v} f_2(\vec{r},\vec{v},t) \right] + \operatorname{div}_v \left[ \langle \dot{\vec{v}} \rangle (\vec{r},\vec{v},t) f_2(\vec{r},\vec{v},t) \right] = 0, \quad (i.4)$$

where

$$f_1(\vec{r},t) = \int_{(\infty)} f_2(\vec{r},\vec{v},t) d^3v, \; N(t) = \int_{(\infty)} f_1(\vec{r},t) d^3r, \quad (i.5)$$

$$f_1(\vec{r},t)\langle \vec{v} \rangle (\vec{r},t) = \int_{(\infty)} \vec{v} f_2(\vec{r},\vec{v},t) d^3v, \; f_2(\vec{r},\vec{v},t)\langle \dot{\vec{v}} \rangle (\vec{r},\vec{v},t) = \int_{(\infty)} \dot{\vec{v}} f_3(\vec{r},\vec{v},\dot{\vec{v}},t) d^3\dot{v}.$$

The vector fields $\langle \vec{v} \rangle (\vec{r},t)$ and $\langle \dot{\vec{v}} \rangle (\vec{r},\vec{v},t)$ correspond to the speed and acceleration of the probability flows. The function $N(t)$ determines the number of particles in the system, which can be non-integer [16]. For a constant number of particles ($N = const$) the value $N$ is used as a normalizing factor when calculating the total probability. The distribution function $f_3(\vec{r},\vec{v},\dot{\vec{v}},t)$ satisfies the third Vlasov equation. Note that the variables $\vec{r},\vec{v},\dot{\vec{v}},\ddot{\vec{v}},...$ are independent kinematic quantities.

For the first Vlasov equation (i.3) the following Hamilton-Jacobi equation is valid [17, 25]:

$$-\hbar \frac{\partial \varphi_1}{\partial t} = \frac{m}{2} |\langle \vec{v} \rangle|^2 + e\chi_1 = H_1, \quad (i.5)$$

$$e\chi_1 \stackrel{det}{=} U_1 + Q_1 + \frac{e^2}{2m}|\vec{A}_1|^2, \quad Q_1 = Q = \frac{\alpha_1}{\beta_1} \frac{\Delta_r |\Psi_1|}{|\Psi_1|} = -\frac{\hbar^2}{2m} \frac{\Delta_r |\Psi_1|}{|\Psi_1|}, \quad (i.6)$$

where $\vec{A}_1$ is the vector potential; $U_1$ is the potential from Schrödinger equation; $\varphi_1$ is the phase of wave function $\Psi_1$; $f_1 = |\Psi_1|^2$; quantity $Q_1$ is the quantum potential from the de Broglie-Bohm theory of the «pilot wave» [26-29]. The quantum potential $Q_1$ allows us to determine the quantum pressure tensor $P^{(q)}_{\mu\lambda}$:

$$-\frac{1}{f_1} \frac{\partial P^{(q)}_{\mu\lambda}}{\partial x^\lambda} = 2\alpha_1^2 \frac{\partial}{\partial x^\mu} \left( \frac{1}{\sqrt{f_1}} \frac{\partial^2 \sqrt{f_1}}{\partial x^\lambda \partial x^\lambda} \right) = 2\alpha_1 \beta_1 \frac{\partial Q_1}{\partial x^\mu}. \quad (i.7)$$

The motion equations are as follows:

$$\frac{d}{dt}\langle\vec{v}\rangle = -\gamma_1\left(\vec{E}_1 + \langle\vec{v}\rangle \times \vec{B}_1\right), \tag{i.8}$$

$$\vec{E}_1 = -\frac{\partial}{\partial t}\vec{A}_1 - \nabla_r \chi_1,$$

or

$$P_{\mu\lambda} = \int\limits_{(\infty)} f_2(\vec{r},\vec{v},t)\left(v_\mu - \langle v_\mu\rangle\right)\left(v_\lambda - \langle v_\lambda\rangle\right)d^3v, \tag{i.9}$$

$$\frac{d_1}{dt}\langle v_\mu\rangle \overset{\text{det}}{=} \left(\frac{\partial}{\partial t} + \langle v_\kappa\rangle\frac{\partial}{\partial x_\kappa}\right)\langle v_\mu\rangle = -\frac{1}{f_1}\frac{\partial P_{\mu\lambda}}{\partial x^\lambda} + \langle\langle\dot{v}_\mu\rangle\rangle. \tag{i.10}$$

The Wigner function evolution is described by the Moyal equation [13]:

$$\frac{\partial W}{\partial t} + \frac{1}{m}(\vec{p},\nabla_r)W - (\nabla_r U, \nabla_p W) = \sum_{l=1}^{+\infty}\frac{(-1)^l(\hbar/2)^{2l}}{(2l+1)!}U\left(\overleftarrow{\nabla}_r,\vec{\nabla}_p\right)^{2l+1}W, \tag{i.11}$$

where potential $U$ is an analytical function.

It was shown [16] that the Moyal equation can be regarded as a special case of the second Vlasov equation (i.4) with the Vlasov-Moyal approximation [16] of the vector acceleration field $\langle\dot{\vec{v}}\rangle(\vec{r},\vec{v},t)$:

$$\langle\dot{v}_\mu\rangle = \sum_{n=0}^{+\infty}\frac{(-1)^{n+1}(\hbar/2)^{2n}}{m^{2n+1}(2n+1)!}\frac{\partial^{2n+1}U_1}{\partial x_\mu^{2n+1}}\frac{1}{f_2}\frac{\partial^{2n}f_2}{\partial v_\mu^{2n}}. \tag{i.12}$$

This implies a direct connection between the Wigner function and the second Vlasov quasi-distribution function $f_2(\vec{r},\vec{v},t)$. However, the Vlasov formalism has its advantages. Firstly, it is based on the first principle – the probability conservation law and does not contain phenomenological constructions such as, for example, the phenomenological Wigner function (i.1) definition. Secondly, it applies to the distribution functions depending on kinematic quantities of all orders $\vec{r},\vec{v},\dot{\vec{v}},\ddot{\vec{v}},...$. And this circumstance can be used to generalize quantum mechanics to the case of higher order kinematic quantities [17].

The well-known Wigner function of a harmonic oscillator with potential $U_1(x) = \frac{m\omega^2 x^2}{2}$ has the following form:

$$f_{2,n}(x,v) = \frac{(-1)^n m}{\pi\hbar}e^{-\frac{m}{\hbar\omega}(v^2+\omega^2 x^2)}L_n\left(\frac{2m}{\hbar\omega}(v^2+\omega^2 x^2)\right), \tag{i.13}$$

where $L_n$ are the Laguerre polynomials. The function $f_{2,n}(x,v)$ is related to the Wigner function from its definition (i.1) as $W_n(x,p) = \frac{1}{m} f_{2,n}\left(x, \frac{p}{m}\right)$. Note that function (i.13) can be written as $f_{2,n}(x,v) = F_n(\varepsilon(x,mv))$:

$$F_n(\varepsilon) = \frac{(-1)^n m}{\pi \hbar} e^{-2\varepsilon} L_n(4\varepsilon), \qquad (i.14)$$

where

$$\varepsilon(x,p) = \frac{1}{\hbar \omega}\left(\frac{p^2}{2m} + \frac{m\omega^2 x^2}{2}\right). \qquad (i.15)$$

The second Vlasov equation (i.4) for a harmonic oscillator takes the form:

$$v\frac{\partial}{\partial x} f_{2,n}(x,v) - \omega^2 x \frac{\partial}{\partial v} f_{2,n}(x,v) = 0, \quad \int_{-\infty}^{+\infty} f_{2,n}(x,v)dv = f_{1,n}(x). \qquad (i.16)$$

Averaging over the velocity space of the Vlasov-Moyal approximation (i.12) will give the classical Vlasov approximation [18] for equation (i.10):

$$\langle\langle \dot{v}_\mu \rangle\rangle = -\frac{1}{m}\frac{\partial U_1}{\partial x_\mu}. \qquad (i.17)$$

It was shown [17] that function (i.13) can be obtained not only from the definition (i.1), but also, directly, from the evolution equation of the quasi-distribution function which can be represented by either the Moyal equation (i.11) or the second Vlasov equation (i.16) with Vlasov-Moyal approximation (i.12). Regardless of the obtaining method, the Wigner function (i.13) allows one to perform quantum averaging in the same way as it is done in classical statistical physics. Thus, in the case of quantum mechanics, one can formally introduce analogs of such quantities as potential and kinetic energy. Despite the seeming formality of such constructions the results obtained can be interpreted in an interesting way.

The present paper has the following structure. In Section 1 we consider the obtaining of explicit expressions for the average (in coordinate / velocity) kinematic quantities corresponding to the «kinetic» and «potential» energy of a quantum harmonic oscillator. In Section 2 it is shown that the average values of the energies $\langle\varepsilon\rangle_v(x)$, $\langle\varepsilon\rangle_x(v)$ have poles located in the regions where the Wigner function takes negative values. The energy poles $\langle\varepsilon\rangle_v(x)$ form infinite potential barriers which number increases along with the growth of the state number $n$ of the quantum system. The conclusion section contains an interpretaion of the main results. The details of mathematical transformations are presented in the Appendix.

## §1 Average kinematic values calculation

Knowing the distribution function $f_{2,n}(x,v)$ (i.13) it is possible to calculate the average values of the energy (i.15) $E = \langle\langle\varepsilon\rangle\rangle = M\varepsilon$ and its standard deviation $\sigma_E = \sqrt{D\varepsilon}$:

$$E_n = \langle\langle\varepsilon\rangle\rangle_n = \int_{-\infty}^{+\infty}\int_{-\infty}^{+\infty} f_{2,n}(x,v)\varepsilon(x,v)dxdv = \hbar\omega\left(n+\frac{1}{2}\right), \quad (1.2)$$

$$\sigma_{E_n}^2 = \int_{-\infty}^{+\infty}\int_{-\infty}^{+\infty} f_{2,n}(x,v)\left[\varepsilon(x,v)-E_n\right]^2 dxdv = \frac{\hbar\omega}{2}. \quad (1.3)$$

For each state the standard deviation energy (1.3) is the same. In this case the distances between the average values $E_n$ of the energy according to the expressions (1.2) and (1.3) are equal $2\sigma_{E_n}$. Thus from the standpoint of quantum mechanics various (continuous spectrum) energy values $\varepsilon$ «exist» in the phase space, but the set of their average values (according to the Wigner distribution function) is countable and coincides with the eigenvalues of the Hamiltonian. The use of phase space allows for a visual interpretation of the relationship between classical and quantum mechanics.

Despite the incorrectness of reasoning about the kinetic and potential energy of a quantum system from a mathematical point of view it is possible to formally calculate these quantities and illustrate a number of interesting relationships between classical and quantum mechanics in phase space.

Let's calculate the average value of the «kinetic» $T = \frac{mv^2}{2}$ and «potential» $U_1 = \frac{m\omega^2 x^2}{2}$ energy of the quantum harmonic oscillator. The averaging can be performed over coordinate $x$, velocity $v$, or over the entire phase space (over both variables $x,v$). Let us find the values $\langle x^2\rangle$, $\langle v^2\rangle$, $\langle\langle x^2\rangle\rangle$, $\langle\langle v^2\rangle\rangle$ for different states $n$ of the system which are described by the distribution function $f_{2,n}(x,v)$ (i.13). Note that due to the symmetry of the distribution function $f_{2,n}(x,v)$, the average values of $\langle x\rangle = \langle\langle x\rangle\rangle = 0$ and $\langle v\rangle = \langle\langle v\rangle\rangle = 0$.

For convenience purposes we introduce designations of $\sigma_x^2 = \frac{\hbar}{2m\omega}$ and $\sigma_v^2 = \frac{\hbar\omega}{2m}$, which correspond to the standard deviation for the ground state ($n=0$) of a harmonic oscillator. The quantities $\sigma_x$, $\sigma_v$ are related to the Heisenberg uncertainty principle

$$\sigma_x \sigma_v = |\alpha| = \frac{\hbar}{2m}, \quad \omega = \frac{\sigma_v}{\sigma_x}. \quad (1.4)$$

Using the quantities $\sigma_x$ and $\sigma_v$ the expression for the energy (i.15) and the representations of the distribution functions (i.5) and (i.13) can be rewritten in the form:

$$2\varepsilon(x,p) = \tilde{\varepsilon}(x,v) = \frac{v^2}{2\sigma_v^2} + \frac{x^2}{2\sigma_x^2}, \quad (1.5)$$

$$f_{1,n}(x) = \frac{1}{2^n n!}\frac{1}{\sqrt{2\pi}\sigma_x} e^{-\frac{x^2}{2\sigma_x^2}} H_n^2\left(\frac{x}{\sqrt{2}\sigma_x}\right), \quad f_{2,n}(x,v) = \frac{(-1)^n}{2\pi\sigma_v\sigma_x} e^{-\frac{v^2}{2\sigma_v^2}-\frac{x^2}{2\sigma_x^2}} L_n\left(2\left(\frac{v^2}{2\sigma_v^2}+\frac{x^2}{2\sigma_x^2}\right)\right).$$

Performing calculations we find $\langle v^2\rangle$ [Appendix A]:

$$\langle v^2 \rangle_{v,n}(x) = \sigma_v^2 \frac{\sum_{k=0}^{n} C_k L_{n-k}\left(\frac{x^2}{\sigma_x^2}\right)}{\sum_{k=0}^{n} \bar{C}_k L_{n-k}\left(\frac{x^2}{\sigma_x^2}\right)}, \tag{1.6}$$

$$C_k = (-1)^k \sum_{s=0}^{k} \frac{1+\eta(s)}{2} \frac{H_{k-s}^2(0) + 2(k-s)H_{k-s-1}^2(0)}{2^{k-s}(k-s)!},$$

$$\bar{C}_k = (-1)^k \sum_{s=0}^{k} \frac{1+\eta(s)}{2} \frac{H_{k-s}^2(0) - 2(k-s)H_{k-s-1}^2(0)}{2^{k-s}(k-s)!},$$

where $\eta(s)$ is the Heaviside function. To calculate the coefficients $\bar{C}_k$ it is convenient to use the properties of the zeros of the Hermite polynomials:

$$H_{2k}^2(0) = \frac{(2k)!(2k)!}{k!k!}, \quad H_{2k+1}^2(0) = 0. \tag{1.7}$$

The index «$v$» in the expression (1.6) indicates the averaging over the space of velocities and the index «$n$» corresponds to the state number of the quantum harmonic oscillator.

Due to the symmetry of the function $f_{2,n}$ with respect to the variables $x$ and $v$ and similarily as with expression (1.6) we can obtain an expression for $\langle x^2 \rangle_{x,n}$ when averaging over the coordinate space $x$ [Appendix A]:

$$\langle x^2 \rangle_{x,n}(v) = \sigma_x^2 \frac{\sum_{k=0}^{n} C_k L_{n-k}\left(\frac{v^2}{\sigma_v^2}\right)}{\sum_{k=0}^{n} \bar{C}_k L_{n-k}\left(\frac{v^2}{\sigma_v^2}\right)}. \tag{1.8}$$

## §2 The Energy function poles distribution in the phase space

Let us analyze the obtained distributions (1.6) and (1.8). A peculiarity of expressions (1.6) and (1.8) is the presence of poles for the coordinate and velocity respectively. Fig. 1 shows the oscillator energy dependence

$$\langle \tilde{\varepsilon} \rangle_{v,n}(x) = \frac{\langle v^2 \rangle_{v,n}(x)}{2\sigma_v^2} + \frac{x^2}{2\sigma_x^2}, \tag{2.1}$$

along the coordinate axis ($n = 0, 1, 2, 3$), a similar graph can be shown along the velocity axis. Without loss of generality the values $\sigma_x$ and $\sigma_v$ in fig. 1 are taken equal to the value of unity. Fig. 1 shows that the ground state ($n = 0$) only has no poles since $\langle v^2 \rangle_{v,0} = \sigma_v^2$ (see expression (1.6)).

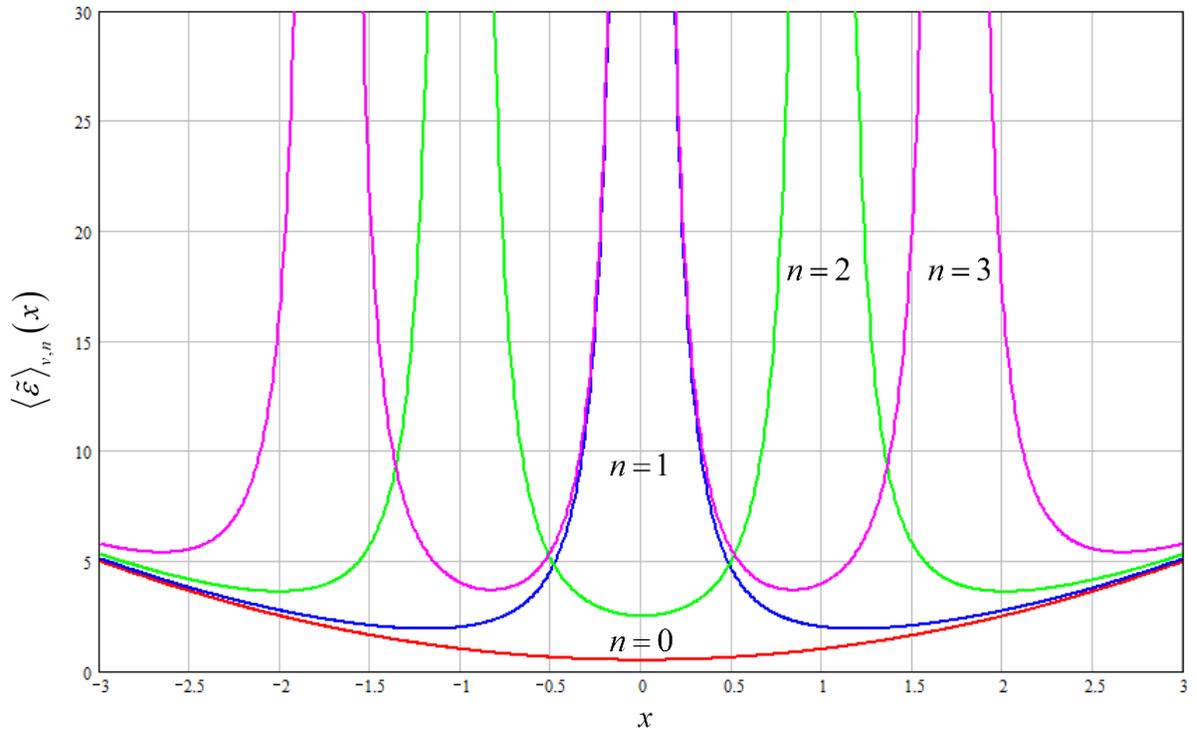

Figure 1. Energy density distribution $\langle \tilde{\varepsilon} \rangle_{v,n}(x)$

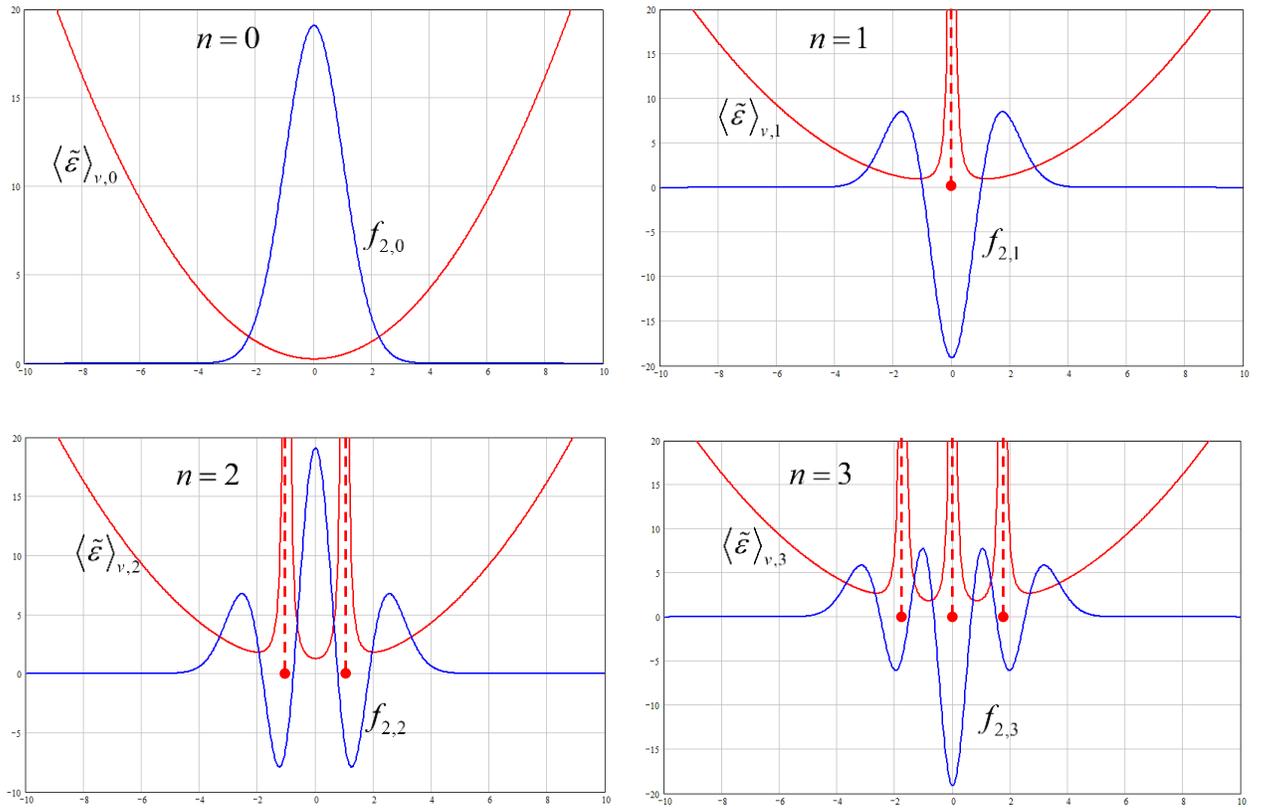

Figure 2. Distribution of energy density $\langle \tilde{\varepsilon} \rangle_{v,n}(x)$ and Wigner function $f_{2,n}(x,0)$

Note that the poles of the kinetic and potential energy (1.6), (1.8) are located at the zeros of the Hermite polynomials (1.7) and are in the domain of negative values of the Wigner function (see fig. 2). In fig. 2 the superposition is shown of the energy plots (2.1) on the corresponding distributions of the Wigner function for the states $n = 0,1,2,3$. Fig. 3 demonstrates a formal overlay of the graphs of the density of probability distribution $f_{1,n}(x)$ and the density of energy distribution $\langle \tilde{\varepsilon} \rangle_{v,n}(x)$ for the states $n = 0,1,2,3$ respectively.

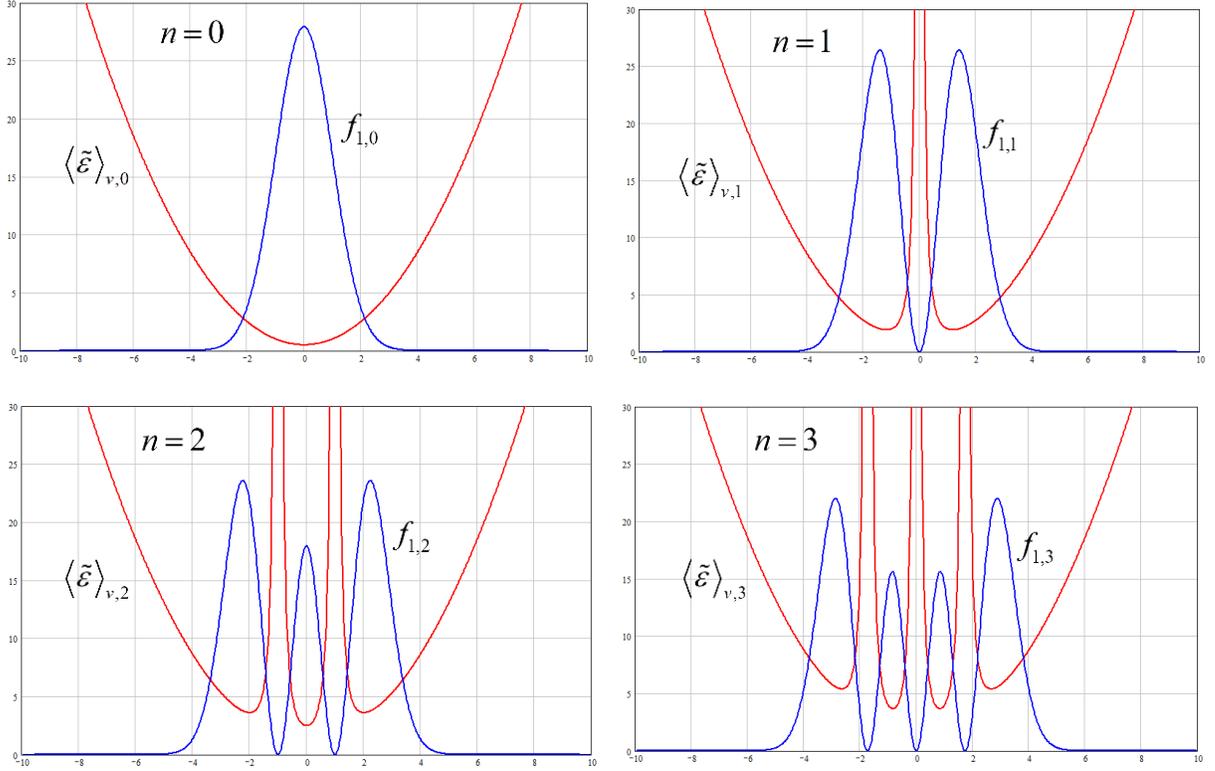

Figure 3. Distribution of energy density $\langle \tilde{\varepsilon} \rangle_{v,n}(x)$ and probability density $f_{1,n}(x)$

Fig. 3 shows that the poles actually «break» the potential well into several potential wells, each of which formally has its own oscillatory process. Indeed, in fig. 2 there are no poles for the ground state ($n=0$). Therefore there is only one initial potential well in which there is a Gaussian probability density distribution $f_{1,0}$. Having $n=1$ the kinetic energy has a pole at zero (see fig. 1, 3) which leads to the presence of an energy «barrier» and the division of the distribution function $f_{1,0}$ into two symmetric distributions relative to zero in the form of a distribution function $f_{1,1}$ (see fig. 3). A similar situation is observed for states with $n = 2,3...$ (see fig. 3).

For a harmonic oscillator the transition from one quantum state with a number $n$ to a state with a number $n+1$ is associated with the presence of negative values in the Wigner function.

It is in the domain of negative values that the kinetic (1.6) and potential energies (1.8) have poles which lead to energy barriers (see fig. 1-3) and formally «create» one more oscillator.

Let us calculate the standard deviations $\langle\langle x^2 \rangle\rangle_n$ and $\langle\langle v^2 \rangle\rangle_n$ [Appendix B]:

$$\left\langle\left\langle v^{2}\right\rangle\right\rangle_{n}=\sigma_{v}^{2}(2n+1), \qquad \left\langle\left\langle x^{2}\right\rangle\right\rangle_{n}=\sigma_{x}^{2}(2n+1). \qquad (2.2)$$

In the expression (2.2) it can be seen that for the ground state ($n=0$) the standard deviations $\sqrt{\left\langle\left\langle v^{2}\right\rangle\right\rangle_{0}}=\sigma_{v}$ and $\sqrt{\left\langle\left\langle x^{2}\right\rangle\right\rangle_{0}}=\sigma_{x}$ or in other words coincide with the values $\sigma_{v}$ and $\sigma_{x}$ introduced formally above (1.4). Consequently, the notation (1.4) has a clear interpretation. As the state number $n$ increases the quantities (2.2) grow. Knowing $\left\langle\left\langle x^{2}\right\rangle\right\rangle_{n}$ and $\left\langle\left\langle v^{2}\right\rangle\right\rangle_{n}$ we can calculate the total average energy of the harmonic oscillator (1.2) in the state $n$:

$$E = T + U_1 = \frac{mv^2}{2} + \frac{m\omega^2 x^2}{2},$$

$$E_n = \frac{m}{2}\sigma_v^2(2n+1) + \frac{m\omega^2}{2}\sigma_x^2(2n+1) = m\left(n+\frac{1}{2}\right)\left(\sigma_v^2 + \omega^2\sigma_x^2\right),$$

$$E_n = \hbar\omega\left(n+\frac{1}{2}\right), \qquad (2.3)$$

where the relations (1.4) are taken into account. The resulting expression (2.3) completely coincides with the expression (1.2). Thus, despite the incorrectness of reasoning about the kinetic and potential energy separately from the standpoint of the Heisenberg uncertainty principle such contradictions are leveled out for quantum mechanics in the phase space.

**Conclusions**
The position of poles of the expressions $\langle\tilde{\varepsilon}\rangle_{v,n}(x)$ / $\langle\tilde{\varepsilon}\rangle_{x,n}(v)$ in the domains where the corresponding Wigner function $f_{2,n}(x,0)$ / $f_{2,n}(x,0)$ has negative values is associated with the presence of zeros for the functions $f_{1,n}(x)=|\Psi_n(x)|^2$ / $\tilde{f}_{1,n}(v)=|\tilde{\Psi}_n(v)|^2$. This statement will remain true for a quantum system with a higher order polynomial potential [20, 21]. In the case of the second degree potential the position of the energy poles $\langle\tilde{\varepsilon}\rangle_{v,n}(x)$ / $\langle\tilde{\varepsilon}\rangle_{x,n}(v)$ is determined by the zeros of the Hermite polynomials (1.5), (1.7).

**Acknowledgements**
This work was supported by the RFBR No. 18-29-10014. This research has been supported by the Interdisciplinary Scientific and Educational School of Moscow University «Photonic and Quantum Technologies. Digital Medicine».

**Appendix A**

From equations (i.6), (i.8), (i.9), (i.10), (i.17) and (i.16) for a harmonic oscillator it follows that

$$\frac{1}{f_1}\frac{\partial P}{\partial x} = \frac{1}{f_1}\int_{-\infty}^{+\infty} v^2 \frac{\partial f_{2,n}}{\partial x}dv = -\frac{1}{m}\frac{\partial U_1}{\partial x} = -\omega^2 x,$$

where in the one-dimensional case (i.9) the notation of $P_{\mu\lambda}$ is changed to $P$.

$$\frac{x}{\sigma_x^2}\int_{-\infty}^{+\infty} v^2 F_n'(\tilde{\varepsilon})dv = -\omega^2 x f_{1,n} = -\omega^2 x \int_{-\infty}^{+\infty} F_n(\tilde{\varepsilon})dv,$$

$$0 = \int_{-\infty}^{+\infty}\left(v^2 F_n'(\tilde{\varepsilon}) + \sigma_x^2 \omega^2 F_n(\tilde{\varepsilon})\right)dv = \int_{-\infty}^{+\infty}\left(v^2 \frac{\partial S_{2,n}}{\partial \tilde{\varepsilon}} + \sigma_x^2 \omega^2\right)F_n(\tilde{\varepsilon})dv = f_{1,n}\left\langle v^2 \frac{\partial S_{2,n}}{\partial \tilde{\varepsilon}} + \sigma_x^2 \omega^2 \right\rangle,$$

$$\left\langle v^2 \frac{\partial S_{2,n}}{\partial \tilde{\varepsilon}} + \sigma_x^2 \omega^2 \right\rangle = 0, \qquad \left\langle v^2 \frac{\partial S_{2,n}}{\partial \tilde{\varepsilon}} \right\rangle = -\sigma_x^2 \omega^2, \tag{A.1}$$

where $S_{2,n} = \mathrm{Ln}\, f_{2,n}$. Rewriting condition (i.7) in the form $P_{\mu\lambda} = -\alpha^2 f_{1,n}\frac{\partial^2 S_{1,n}}{\partial x^\mu \partial x^\lambda}$, $S_{1,n} = \mathrm{Ln}\, f_{1,n}$ we obtain

$$\frac{1}{f_{1,n}}\int_{-\infty}^{+\infty} v^2 f_{2,n}(x,v)dv = -\alpha^2 \frac{\partial^2 S_{1,n}}{\partial x^2}, \qquad \left\langle v^2 \right\rangle = -\alpha^2 \frac{\partial^2 S_{1,n}}{\partial x^2}. \tag{A.2}$$

Substitute the distribution functions (i.14) into expressions (A.1) and (A.2). Let us start with the expression (A.1).

$$S_{2,n} = \mathrm{Ln}\, F_n(\tilde{\varepsilon}) = \mathrm{Ln}\, B_n - \tilde{\varepsilon} + \mathrm{Ln}\, L_n(2\tilde{\varepsilon}), \quad \frac{\partial S_{2,n}}{\partial \tilde{\varepsilon}} = -1 + 2\frac{L_n'(2\tilde{\varepsilon})}{L_n(2\tilde{\varepsilon})}, \tag{A.3}$$

where $B_n = \frac{(-1)^n}{2\pi\sigma_v\sigma_x}$. Averaging expression (A.3) using (A.1) we obtain

$$\sigma_x^2\omega^2 = \frac{B_n}{f_{1,n}}\int_{-\infty}^{+\infty} v^2 e^{-\tilde{\varepsilon}}\left(L_n(2\tilde{\varepsilon}) + 2L_{n-1}^{(1)}(2\tilde{\varepsilon})\right)dv, \tag{A.4}$$

where we take into account that $L_n' = L_{n-1}' - L_{n-1}$, $L_s^{(\mu+1)} = \sum_{k=0}^{s} L_k^{(\mu)}$. Considering the expression $L_n^{(\mu)}(x) = L_n^{(\mu+1)}(x) - L_{n-1}^{(\mu+1)}(x)$ which at $\mu = 0$ will be $L_n(x) = L_n^{(1)}(x) - L_{n-1}^{(1)}(x)$ the expression (A.4) will take the form:

$$\frac{\sigma_v^2}{B_n}f_{1,n} = \int_{-\infty}^{+\infty} v^2 e^{-\tilde{\varepsilon}}\left(L_n^{(1)}(2\tilde{\varepsilon}) + L_{n-1}^{(1)}(2\tilde{\varepsilon})\right)dv. \tag{A.5}$$

The generalized Laguerre polynomials satisfy the relations

$$L_n^{(1)}(2\tilde{\varepsilon}) = \sum_{k=0}^{n} L_k\left(\frac{v^2}{\sigma_v^2}\right) L_{n-k}\left(\frac{x^2}{\sigma_x^2}\right),$$

$$L_{n-1}^{(1)}(2\tilde{\varepsilon}) = \sum_{k=0}^{n-1} L_k\left(\frac{v^2}{\sigma_v^2}\right) L_{n-1-k}\left(\frac{x^2}{\sigma_x^2}\right). \tag{A.6}$$

Substituting (A.6) into (A.4) we obtain

$$(-1)^n \frac{\sqrt{\pi}}{2^{n+1} n!} H_n^2\left(\frac{x}{\sqrt{2}\sigma_x}\right) = \sum_{k=1}^{n} L_{n-k}\left(\frac{x^2}{\sigma_x^2}\right)(J_k + J_{k-1}) + L_n\left(\frac{x^2}{\sigma_x^2}\right) J_0, \tag{A.7}$$

where

$$J_k = \int_{-\infty}^{+\infty} \tau^2 e^{-\tau^2} L_k(2\tau^2) d\tau, \quad \tau = \frac{v}{\sqrt{2}\sigma_v}. \tag{A.8}$$

Expression (A.7) allows representing the square of the Hermite polynomials $H_n^2$ in terms of the Laguerre polynomials $L_k$. Let us calculate the integral (A.8).

$$J_k = \frac{1}{2} \int_{-\infty}^{+\infty} e^{-\tau^2} L_k(2\tau^2) d\tau - 2J_{k-1} + 2\int_{-\infty}^{+\infty} \tau^2 e^{-\tau^2} L'_{k-1}(2\tau^2) d\tau. \tag{A.9}$$

Considering that $\int_{-\infty}^{+\infty} e^{-\tau^2} L_k(2\tau^2) d\tau = (-1)^k \frac{\sqrt{\pi}}{2^k k!} H_k^2(0)$ we get

$$J_k = (-1)^k \frac{\sqrt{\pi}}{2^{k+1} k!} H_k^2(0) - 2J_{k-1} + 2\int_{-\infty}^{+\infty} \tau^2 e^{-\tau^2} L'_{k-1}(2\tau^2) d\tau,$$

hence,

$$J_{k-1} = (-1)^{k-1} \frac{\sqrt{\pi}}{2^k (k-1)!} H_{k-1}^2(0) - 2J_{k-2} + 2\int_{-\infty}^{+\infty} \tau^2 e^{-\tau^2} L'_{k-2}(2\tau^2) d\tau. \tag{A.10}$$

Substituting (A.10) into expression (A.9) we obtain

$$J_k = (-1)^k \frac{\sqrt{\pi}}{2^{k+1} k!} H_k^2(0) + (-1)^k \frac{\sqrt{\pi}}{2^{k-1}(k-1)!} H_{k-1}^2(0) + 2J_{k-2} - 2\int_{-\infty}^{+\infty} \tau^2 e^{-\tau^2} L'_{k-2}(2\tau^2) d\tau. \tag{A.11}$$

Then let us carry out a similar substitution procedure for the integrals $J_{k-2}$ and $J_{k-3}$:

$$J_k = (-1)^k \frac{\sqrt{\pi}}{2^{k+1}k!} H_k^2(0) + (-1)^k \frac{\sqrt{\pi}}{2^{k-1}(k-1)!} H_{k-1}^2(0) + (-1)^k \frac{\sqrt{\pi}}{2^{k-2}(k-2)!} H_{k-2}^2(0) - $$
$$-2J_{k-3} + 2\int_{-\infty}^{+\infty} \tau^2 e^{-\tau^2} L'_{k-3}(2\tau^2) d\tau.$$
(A.12)

$$J_k = (-1)^k \frac{\sqrt{\pi}}{2^{k+1}k!} H_k^2(0) + (-1)^k \frac{\sqrt{\pi}}{2^{k-1}(k-1)!} H_{k-1}^2(0) + (-1)^k \frac{\sqrt{\pi}}{2^{k-2}(k-2)!} H_{k-2}^2(0) + $$
$$+(-1)^k \frac{\sqrt{\pi}}{2^{k-3}(k-3)!} H_{k-3}^2(0) + 2J_{k-4} - 2\int_{-\infty}^{+\infty} \tau^2 e^{-\tau^2} L'_{k-4}(2\tau^2) d\tau.$$
(A.13)

The expression for $J_0$ has the form

$$J_0 = \int_{-\infty}^{+\infty} \tau^2 e^{-\tau^2} L_0(2\tau^2) d\tau = \frac{1}{2\sqrt{2}\sigma_v^3} \int_{-\infty}^{+\infty} v^2 e^{-\frac{v^2}{2\sigma_v^2}} dv = \frac{\sqrt{2\pi}}{2\sqrt{2}} = \frac{\sqrt{\pi}}{2}.$$
(A.14)

Let us consider the even ($k=2m$) and odd ($k=2m+1$) values for the expression $J_k$. Proceeding with the iterative procedure the expression (A.13) for $J_{2m}$ takes the form:

$$J_{2m} = \frac{\sqrt{\pi}}{2^{2m+1}(2m)!} H_{2m}^2(0) + \sqrt{\pi} \sum_{s=1}^{2m} \frac{H_{2m-s}^2(0)}{2^{2m-s}(2m-s)!}.$$
(A.15)

Similarly, for $J_{2m+1}$ we obtain

$$J_{2m+1} = -\frac{\sqrt{\pi}}{2^{2m+2}(2m+1)!} H_{2m+1}^2(0) - \sqrt{\pi} \sum_{q=1}^{2m+1} \frac{H_{2m+1-q}^2(0)}{2^{2m+1-q}(2m+1-q)!}.$$
(A.16)

Comparing (A.15) and (A.16) we obtain a general expression for $J_k$

$$J_k = (-1)^k \sqrt{\pi} \left\{ \frac{1}{2^{k+1}k!} H_k^2(0) + \sum_{s=1}^{k} \frac{H_{k-s}^2(0)}{2^{k-s}(k-s)!} \right\}.$$
(A.17)

To transform expression (A.7) let us calculate the sum $J_k + J_{k-1}$ using (A.17)

$$J_k + J_{k-1} = (-1)^k \sqrt{\pi} \left\{ \frac{H_k^2(0) - 2kH_{k-1}^2(0)}{2^{k+1}k!} + \sum_{s=1}^{k} \frac{H_{k-s}^2(0) - 2(k-s)H_{k-s-1}^2(0)}{2^{k-s}(k-s)!} \right\}.$$
(A.18)

Substituting (A.18) into (A.7) we get

$$\frac{(-1)^n}{2^{n+1}n!}H_n^2\left(\frac{x}{\sqrt{2}\sigma_x}\right)=\frac{1}{2}L_n\left(\frac{x^2}{\sigma_x^2}\right)+$$
$$+\sum_{k=1}^{n}(-1)^k L_{n-k}\left(\frac{x^2}{\sigma_x^2}\right)\left\{\frac{H_k^2(0)-2kH_{k-1}^2(0)}{2^{k+1}k!}+\sum_{s=1}^{k}\frac{H_{k-s}^2(0)-2(k-s)H_{k-s-1}^2(0)}{2^{k-s}(k-s)!}\right\}.$$
(A.19)

Expression (A.19) can be rewritten in a compact form using the Heaviside function

$$\eta(s)=\begin{cases}0, & s=0,\\ 1, & s>0.\end{cases}\qquad \frac{1+\eta(s)}{2}=\begin{cases}\frac{1}{2}, & s=0,\\ 1, & s>0.\end{cases}$$
(A.20)

Using (A.20) the expression (A.19) takes the following form

$$\frac{(-1)^n}{2^{n+1}n!}H_n^2\left(\frac{x}{\sqrt{2}\sigma_x}\right)=\sum_{k=0}^{n}\bar{C}_k L_{n-k}\left(\frac{x^2}{\sigma_x^2}\right),$$
(A.21)

where

$$\bar{C}_k=(-1)^k\sum_{s=0}^{k}\frac{1+\eta(s)}{2}\frac{H_{k-s}^2(0)-2(k-s)H_{k-s-1}^2(0)}{2^{k-s}(k-s)!}.$$

Now let us consider the expression (A.2). First of all we calculate the expression $\frac{\partial^2 S_1}{\partial x^2}$ in (A.2)

$$\frac{\partial^2 S_1}{\partial x^2}=-\frac{1}{\sigma_x^2}\left(1-\frac{H_n''}{H_n}+\left(\frac{H_n'}{H_n}\right)^2\right).$$
(A.22)

And find $\langle v^2\rangle$

$$\frac{(-1)^n\sqrt{\pi}}{2^{n+1}n!\sigma_v^2}H_n^2\left(\frac{x}{\sqrt{2}\sigma_x}\right)\langle v^2\rangle=\sum_{k=0}^{n}L_{n-k}\left(\frac{x^2}{\sigma_x^2}\right)J_k-\sum_{k=0}^{n-1}L_{n-1-k}\left(\frac{x^2}{\sigma_x^2}\right)J_k,$$
(A.23)

since $\int_{-\infty}^{+\infty}v^2 e^{-\frac{v^2}{2\sigma_v^2}}L_k\left(\frac{v^2}{\sigma_v^2}\right)dv=2\sqrt{2}\sigma_v^3 J_k$. Substituting (A.17) into (A.23) we obtain

$$\frac{(-1)^n}{2^{n+1}n!\sigma_v^2}H_n^2\left(\frac{x}{\sqrt{2}\sigma_x}\right)\langle v^2\rangle=\sum_{k=0}^{n}C_k L_{n-k}\left(\frac{x^2}{\sigma_x^2}\right),$$
(A.24)

where

$$C_k = (-1)^k \sum_{s=0}^{k} \frac{1+\eta(s)}{2} \frac{H^2_{k-s}(0) + 2(k-s)H^2_{k-s-1}(0)}{2^{k-s}(k-s)!}.$$

**Appendix B**

Let us calculate

$$\langle\langle v^2 \rangle\rangle_n = \int_{-\infty}^{+\infty} f_{1,n}(x) \langle v^2 \rangle_{v,n}(x) dx = \frac{\alpha^2}{\sigma_x^2} - \frac{\alpha^2}{\sigma_x^2} \int_{-\infty}^{+\infty} f_{1,n}(x) \frac{H_n''}{H_n} dx + \frac{\alpha^2}{\sigma_x^2} \int_{-\infty}^{+\infty} f_{1,n}(x) \left(\frac{H_n'}{H_n}\right)^2 dx,$$

$$\langle\langle v^2 \rangle\rangle_n = \frac{\alpha^2}{\sigma_x^2} - \frac{\alpha^2}{\sigma_x^2} \frac{1}{2^n n!} \frac{\sqrt{2}\sigma_x}{\sqrt{2\pi}\sigma_x} \int_{-\infty}^{+\infty} e^{-y^2} H_n(y) H_n''(y) dy +$$

$$+ \frac{\alpha^2}{\sigma_x^2} \frac{1}{2^n n!} \frac{\sqrt{2}\sigma_x}{\sqrt{2\pi}\sigma_x} \int_{-\infty}^{+\infty} e^{-y^2} H_n'(y) H_n'(y) dy,$$

taking into account the differentiation formula for the Hermite polynomials $H_n'(y) = 2nH_{n-1}(y)$ and the orthogonality condition we obtain the following expression

$$\langle\langle v^2 \rangle\rangle_n = \frac{\alpha^2}{\sigma_x^2} + \frac{\alpha^2}{\sigma_x^2} \frac{1}{2^n n!} \frac{4n^2}{\sqrt{\pi}} \sqrt{\pi} 2^{n-1}(n-1)! = \frac{\alpha^2}{\sigma_x^2}(2n+1),$$

$$\langle\langle v^2 \rangle\rangle_n = \sigma_v^2(2n+1), \tag{B.1}$$

where the relation (1.4) is taken into account. By virtue of the symmetry of expressions (1.6) and (1.8) we can rewrite the expression $\langle\langle x^2 \rangle\rangle_n$ in a similar way

$$\langle\langle x^2 \rangle\rangle_n = \sigma_x^2(2n+1). \tag{B.2}$$